\documentstyle [12pt]{article}

\def\K{{\cal K}}

\newcommand{\lbl}[1]{\label{eq: #1}}
\newcommand{\rf}[1]{\ref{eq: #1}}

\def\R{{\rm I\hspace{-.15em}R}}

\def\b{\begin{equation}}
\def\e{\end{equation}}
\def\bd{\begin{displaystyle}}
\def\ed{\end{displaystyle}}
\def\ba{\begin{array}}
\def\ea{\end{array}}

\def\bee{\begin{enumerate}}
\def\eee{\end{enumerate}}

\def\bes{\begin{eqnarray*}}
\def\ees{\end{eqnarray*}}
\def\be{\begin{eqnarray}}
\def\ee{\end{eqnarray}}

\begin{document}

\title{Auxiliary ``massless'' spin-2 field in de Sitter universe}
\author{ H. Pejhan$^{1}$, M.R. Tanhayi$^{2}$\thanks{e-mail:
m$_{-}$tanhayi@iauctb.ac.ir} and M.V. Takook$^{3}$\thanks{e-mail:
takook@razi.ac.ir} }
\date{\today}
\maketitle \centerline{\it $^1$Department of Physics, Science and
Research Branch, Islamic Azad University, Tehran, Iran }
\centerline{\it $^2$ Department of Physics, Islamic Azad
University, Central Tehran Branch, Tehran, Iran} \centerline{\it
$^3$Department of Physics, Razi University, Kermanshah, Iran}

 \vspace{15pt}

\begin{abstract}
For the tensor field of rank-2 there are two unitary irreducible
representation (UIR) in de Sitter (dS) space denoted by
$\Pi^{\pm}_{2,2}$ and $\Pi^{\pm}_{2,1}$ \cite{9}. In the flat
limit only the $\Pi^{\pm}_{2,2}$ coincides to the UIR of
Poincar\'{e} group, the second one becomes important in the study
of conformal gravity. In the pervious work, Dirac's six-cone
formalism has been utilized to obtain conformally invariant (CI)
field equation for the ``massless'' spin-2 field in dS space
\cite{1}. This equation results in a field which transformed
according to $\Pi^{\pm}_{2,1}$, we name this field the auxiliary
field. In this paper this auxiliary field is considered and also
related two-point function is calculated as a product of a
polarization tensor and ``massless'' conformally coupled scalar
field. This two-point function is de Sitter invariant.

\end{abstract}

\vspace{0.5cm} {\it PACS numbers}: 04.62.+v, 03.70+k, 11.10.Cd,
98.80.H \vspace{0.5cm}

\newpage
\section{Introduction}

Recent astrophysical data indicate that our universe might
currently be in a dS phase. Quantum field theory in dS space-time
has evolved as an exceedingly important subject, studied by many
authors in the course of past decade. The importance of dS space
has been primarily ignited by the study of the inflationary model
of the universe and quantum gravity. The importance of the
massless spin-2 field in the dS space is due to the fact that it
plays the central role in quantum gravity and quantum cosmology in
the linear approximation. In dS space, mass is not an invariant
parameter for the set of observable transformations under the dS
group $SO(1,4)$. Concept of light-cone propagation, however, does
exist and leads to the conformal invariance. ``Massless" is used
in reference to propagation on the dS light-cone (conformal
invariance). The term ``massive'' is refereed to fields that in
their Minkowskian limit (zero curvature) reduce to massive
Minkowskian fields \cite{2}.

It has been shown that the massive and massless conformally
coupled scalar fields in dS space correspond to the principal and
complementary series representations of dS group, respectively
\cite{3}. The massive vector field in dS space has been associated
with the principal series, whereas massless field corresponds to
the lowest representation of the vector discrete series
representation in dS group \cite{4}. The massive and massless
spin-2 fields in dS space have been also associated with the
principal series and the lowest representation of the rank-2
tensor discrete series of dS group, respectively \cite{5,6,7}. \\
It has been shown that CI wave equation for the tensor field of
rank two must obey following fourth order derivative equation
\cite{1}
$$ (Q_2+6)(Q_2+4)\K_{\alpha\beta}=0, $$ where $Q_2$ is the Casimir operator of the dS group. This equation can be considered as two equation of order 2
$$ (Q_2+6)\K_{\alpha\beta}=0, \,\,\,\,\mbox{transforms according to}\,\,
\Pi^{\pm}_{2,2},$$
$$(Q_2+4)\K_{\alpha\beta}=0, \,\,\,\,\mbox{transforms according to}\,\,
\Pi^{\pm}_{2,1}.$$ In the flat limit $(H\rightarrow 0)$ only
$\Pi^{\pm}_{2,2}$  coincides to the UIR of the Poincar\'{e} group,
however both representations become important in studding the CI
field equation in dS space. The representation $\Pi^{\pm}_{2,2}$
has been studied previously, in this paper we study the fields
which transform according to $\Pi^{\pm}_{2,1}$. Since there is no
flat limit to these fields so they are not detectable locally, and
such fields named as auxiliary fields.
\\The organization of this paper and its brief outlook are as
follows. Section $2$ is devoted to a brief review of the dS
massless spin-2 field equations in the ambient space and aslo
Dirac's manifestly covariant formalism has been reviewed. In
Section $3$ we study the auxiliary CI field equations and their
solutions. We show that this field can be written in terms of a
polarization tensor and a massless conformally coupled scalar
field as follows
$$ \K_{\alpha\beta}(x)= {\cal D}_{\alpha\beta}(x,\partial)\phi (x). $$
In section $4$ we calculate two-point function ${\cal
W}_{\alpha\beta \alpha'\beta'}(x,x')$ in 5 dimensional flat space
(ambient space notation) for the field equation. Finally a brief
conclusion and an outlook for further investigation has been
presented. We have supplied some useful mathematical details of
calculations in the appendices.

\setcounter{equation}{0}

\section{de Sitter space and Dirac's six-cone formalism}
 \textsl{\textbf{I) de Sitter space:}} The
dS space-time can be defined by the one-sheeted four-dimensional
hyperboloid:
\begin{equation}
X_H=\{x \in \R^5 ;x^2=\eta_{\alpha\beta} x^\alpha x^\beta
=-H^{-2}=-\frac{3}{\Lambda}\},\;\;\alpha,\beta=0,1,2,3,4,
\end{equation} where $\eta_{\alpha\beta}=$ diag$(1,-1,-1,-1,-1)$
and $H$, $\Lambda$ are the Hobble parameter and cosmological
constant respectively. The dS metric is
$$ds^2=\eta_{\alpha\beta}dx^{\alpha}dx^{\beta}=g_{\mu\nu}^{dS}dX^{\mu}dX^{\nu},\;\;\mu,\nu=0,1,2,3,$$
where the $X^\mu$'s are  $4$ space-time intrinsic coordinates of
the dS hyperboloid. Any geometrical object in this space can be
written in terms of the four local coordinates $X^\mu$
(intrinsics) or in terms of the five global coordinates $x^\alpha$
(ambient space). \\The wave equation for massless tensor fields
$h_{\mu\nu}(X)$, propagating in de Sitter space is
\cite{hiko1,14}: $$ (\Box_H+2H^2)h_{\mu\nu}-(\Box_H-H^2)
g^{dS}_{\mu\nu}h'-{\cal S} \nabla_{\mu}\nabla\cdot h_{\nu}$$
\b\lbl{field-h} +g^{dS}_{\mu\nu}\nabla\cdot\nabla\cdot h
+\nabla_{\mu}\nabla_{\nu}h'=0,\e where $\Box_H$ is the
Laplace-Beltrami operator in dS space, $h'=h_{\mu}^{\mu}$ and
${\cal S}(\alpha\beta)= (\alpha\beta+\beta\alpha)$. $\nabla^\nu$
is the covariant derivative in dS space. The field equation
(\rf{field-h}) is invariant under the following gauge
transformation \b h_{\mu\nu} \longrightarrow
h_{\mu\nu}^{gt}=h_{\mu\nu}+{\cal S}\nabla_{\mu}\chi_{\nu},\e where
$\chi_{\nu}$ is an arbitrary vector field. A general family of
gauge conditions can be chosen \b\lbl{gaugefix1}
\nabla^{\mu}h_{\mu \nu}=\zeta \nabla_{\nu}h',\e where $\zeta$ is
an arbitrary constant. The choice of $\zeta=\frac{1}{2}$ makes the
relation between field and group representation thoroughly
apparent in the ambient space \cite{chdu}. In the following,
ambient space notations is used; in ambient space, the
relationship with UIRs of the dS group becomes straightforward
because the Casimir operators are easy to identify. The tensor
field $\K_{\alpha\beta}$ defined by $\R^5$-variables $x^{\alpha}$
in de Sitter space-time. Useful relations between ambient and
intrinsic coordinates are listed in the appendix. The kinematical
group of the dS space is the $10$-parameter group $SO_0(1,4)$
(connected component of the identity in $SO(1,4)$ ), which is one
of the two possible deformations of the Poincar\'e group. There
are two Casimir operators \begin{equation}
Q^{(1)}_2=-\frac{1}{2}L^{\alpha\beta}L_{\alpha\beta},\;\;\;\;\
Q^{(2)}_2=-W_{\alpha}W^{\alpha},\end{equation} where
\begin{equation}W_{\alpha}=-\frac{1}{8}\epsilon_{\alpha\beta\gamma\sigma\eta}L^{\beta\gamma}L^{\sigma\eta},\end{equation}
with 10 infinitesimal generators
$L_{\alpha\beta}=M_{\alpha\beta}+S_{\alpha\beta}.$ The subscript
$2$ in $Q^{(1)}_2$, $Q^{(2)}_2$ reminds that the carrier space is
constituted by second rank tensors. The orbital part
$M_{\alpha\beta}$, and the action of the spinorial part
$S_{\alpha\beta}$ on a tensor field $\K$ defined on the ambient
space read respectively \cite{8}
$$M_{\alpha\beta}=-i(x_{\alpha}\partial_{\beta}-x_{\beta}\partial_{\alpha})=
-i(x_{\alpha}\bar\partial_{\beta}-x_{\beta}\bar\partial_{\alpha}),$$
\begin{equation}S_{\alpha\beta}\K_{\gamma\delta}=-i(\eta_{\alpha\gamma}\K_{\beta\delta}-\eta_{\beta\gamma} \K_{\alpha\delta} +
 \eta_{\alpha\delta}\K_{\beta\gamma}-\eta_{\beta\delta}
 \K_{\alpha\gamma}).\end{equation} The symbol $\epsilon_{\alpha\beta\gamma\sigma\eta}$ holds
for the usual antisymmetric tensor. $\bar\partial_{\alpha}$ is the
tangential (or transverse) derivative on dS space defined by $$
\bar\partial_{\alpha}=\theta_{\alpha\beta}\partial^{\beta}=\partial_{\alpha}+H^2x_{\alpha}x\cdot\partial,\,\,\,\mbox{with},\,\,x\cdot\bar\partial=0\,,$$
and $\theta_{\alpha\beta}$ is the transverse projector
($\theta_{\alpha\beta}=\eta_{\alpha\beta}+H^2x_{\alpha}x_{\beta}\,$).

The operator $Q^{(1)}_2$ commutes with the action of the group
generators and, as a consequence, it is constant in each UIR. Thus
the eigenvalues of $Q^{(1)}_2$ can be used to classify UIRs {\it
i.e.,}
\begin{equation}
(Q^{(1)}_2-\langle Q^{(1)}_2\rangle){\cal K}(x)=0.
\end{equation}
Following Dixmier \cite{9}, we get a classification scheme using a
pair $(p,q)$ of parameters involved in the following possible
spectral values of the Casimir operators :
\begin{equation}
Q^{(1)}_{p}=\left(-p(p+1)-(q+1)(q-2)\right)I_d ,\qquad\quad
Q^{(2)}_{p}=\left(-p(p+1)q(q-1)\right)I_d\,.
\end{equation}
Three types of scalar, tensorial or spinorial UIRs are
distinguished for $SO(1,4)$ according to the range of values of
the parameters $q$ and $p$ \cite{9,10}, namely: the principal, the
complementary and the discrete series. The flat limit indicates
that for the principal and the complementary series the value of
$p$ bears meaning of spin. For the discrete series case, the only
representation which has a physically meaningful Minkowskian
counterpart is $p=q$ case. Mathematical details of the group
contraction and the physical principles underlying the
relationship between dS and Poincar\'e groups can be found in
Ref.s \cite{11} and \cite{12} respectively. The spin-$2$ tensor
representations relevant to the present work are as follows:
\begin{itemize}
\item[i)] The UIRs $U^{2,\nu}$ in the principal series where
$p=s=2$ and $q=\frac{1}{2 }+i\nu$ correspond to the Casimir
spectral values:
\begin{equation}
\langle Q_2^{\nu}\rangle=\nu^2-\frac{15}{4},\;\;\nu \in \R,
\end{equation}
note that $U^{2,\nu}$ and $U^{2,-\nu}$ are equivalent. \item[ ii)]
The UIRs $V^{2,q}$ in the complementary series where $p=s=2$ and
$q-q^2=\mu,$ correspond to
\begin{equation}
\langle  Q_2^{\mu}\rangle=q-q^2-4\equiv
\mu-4,\;\;\;0<\mu<\frac{1}{4}\,.
\end{equation}
\item[iii)] The UIRs $\Pi^{\pm}_{2,q}$ in the discrete series
where $p=s=2$ correspond to
\begin{equation}
\langle Q_2^{(1)}\rangle=-4, \;\;q=1\; (\Pi^{\pm}_{2,1});
\;\;\langle Q_2^{(2)}\rangle=-6,\;\; q=2 \;(\Pi^{\pm}_{2,2}),
\end{equation}
the sign $\pm$ stands for the helicity of the representation.
\end{itemize} The action of
the Casimir operators $Q_1 $ and $ Q_2$ can be written in the more
explicit form \b Q_1
K_{\alpha}=(Q_0-2)K_{\alpha}+2x_{\alpha}\partial . K
-2\partial_{\alpha} x\cdot K, \e \b Q_2 {\cal K}_{\alpha
\beta}=(Q_0-6){\cal K}_{\alpha \beta}+2{\cal
S}x_{\alpha}\partial\cdot {\cal K}_{\beta}-2{\cal
S}\partial_{\alpha}x\cdot {\cal
K}_{\beta}+2\eta_{\alpha\beta}{\K}',\, \e where $\cal K'$ is the
trace of ${\cal{K}}_{{\alpha}{\beta}}$ and
$Q_0=\frac{1}{2}M_{\alpha\beta}M^{\alpha\beta}$.\vspace{1.5mm}

\textsl{\textbf{II) Dirac's six-cone formalism:}} Dirac's six-cone
formalism originally was defined by Dirac to obtain CI wave
equation \cite{01}. This formalism developed by Mack and Salam
\cite{02} and many others. Dirac's six-cone or Dirac's projection
cone is defined by
\begin{equation} u^{2} \equiv u^{2}_{0}-\vec
u^{2}+u^{2}_{5}=\eta^{ab}u_au_b=0, \;\;\
\eta^{ab}=diag(1,-1,-1,-1,-1,1),\end{equation} where $ \;u_{a} \in
\R^{6},$ and  $ \vec u \equiv(u_{1},u_{2},u_{3},u_{4})$. Reduction
to four dimensional (physical space-time) is achieved by
projection, that is by fixing the degrees of homogeneity of all
fields. Wave equations, subsidiary conditions, etc., must be
expressed in terms of operators that are defined intrinsically on
the cone. These are well-defined operators that map tensor fields
on tensor fields with the same rank on cone $u^2=0$ \cite{13,14}.
It is important to note that on the cone $u^2=0$, the second order
Casimir operator of conformal group, ${\cal Q}_2 $, is not a
suitable operator to obtain CI wave equations. Because for a
symmetric tensor field of rank-2, we have \cite{2,14,15}: $$ {\cal
Q}_2\Psi^{cd}=\frac{1}{2} L_{ab}L^{ab}\Psi^{cd}=\left(-u^2
\partial^2+\hat{N_5}(\hat{N_5}+4)+8\right)\Psi^{cd},$$ where $\hat{N_5}$ is
the conformal-degree operator defined by $ \hat{N_5}\equiv
u^{a}\partial_{a}.$\\  On the cone this operator reduces to a
constant, {\it i.e.} $\hat{N}_5(\hat{N}_5+4)+8$. It is clear that
this operator cannot lead to the wave equations on the cone. The
well-defined operators exist only in exceptional cases. For tensor
fields of degree $ -1,0,1,...$, the intrinsic wave operators are $
\partial^2, (\partial^2)^2, (\partial^2)^3,...$ respectively
\cite{14}. Thus the following CI system of equations, on the cone,
has been used \cite{13}: \b\label{di} \left\{ \ba{rcl}
(\partial_a\partial^a)^n \Psi&=&0,\\
\hat{N_5}\Psi&=&(n-2)\Psi.\ea\right. \e where $\Psi$ is a tensor
field of a definite rank and of a definite symmetry.

Other CI conditions can be added to the above system in order to
restrict the space of the solutions. The conditions \emph{i}-
transversality:  $u_a\Psi^{ab...}=0 ,$ \emph{ii}- tracelessness:
$\Psi_{ab...}^a=0 $ and \emph{iii}-
divergencelessness: $Grad_a\Psi^{ab...}=0 $ are introduced to achieve the above goal.\\
The operator $Grad_a$ unlike $\partial_{a}$ is intrinsic on the
cone, and is defined by \cite{14}:\b Grad_a\equiv
u_a\partial_b\partial^b-(2\hat{N_5}+4)\partial_a \,.\e

In order to project the coordinates on the cone $u^2=0$, to the
$1+4$ dS space we chose the following relation: \b \left\{
\ba{rcl}
x^{\alpha}&=&(Hu^5)^{-1}u^\alpha,\\
x^5&=&Hu^5.\ea\right.\e Note that $x^5$ becomes superfluous when
we deal with the projective cone. It is easy to show that various
intrinsic operators introduced previously now read as:
\begin{enumerate}

\item{ the conformal-degree operator $(\hat{N_5})$} \b \hat{N_5}=
 x_5\frac{\partial}{\partial x_5},\e
\item{the conformal gradient $(Grad_{\alpha})$}  \b Grad_{\alpha}=
-x_{5}^{-1}
\{H^2x_{\alpha}[Q_{0}-\hat{N_5}(\hat{N_5}-1)]+2\bar{\partial}_{\alpha}(\hat{N_5}+1)\},\e
\item{and the powers of d'Alembertian $(\partial_a\partial^a)^n$},
which acts intrinsically on field of conformal degree $(n-2)$, \b
(\partial_{a}
\partial^{a})^n=-H^{2n}x_{5}^{-2n} \prod_{j=1}^{n}[Q_{0}+(j+1)(j-2)]\,. \e
\end{enumerate} In the next section we consider CI field equation
with $n=1$ in $(2.16)$ and its possible solution.

\setcounter{equation}{0}
\section{Auxiliary CI equation and its solution}
Dirac's six-cone formalism provides us with the opportunity to
obtain CI wave equations for scalar, vector and tensor fields. It
is shown \cite{13} that for the scalar and vector field, the
simplest CI system of equations is obtained through $n=1$ in
(\ref{di}), {\it i.e.} the field with conformal degree $-1$.
Resulting equations are the UIRs of $SO(1,4)$. In the flat limit
$(H\rightarrow 0)$ the CI equation for the vector field reduces
exactly to the Maxwell equation \cite{6} and CI scalar field in
this limit leads to the standard CI wave equation in Minkowskian
space. For symmetric tensor field of rank-2, the CI system
(\ref{di}) for $n=1$ leads to \cite{1} (for simplicity from now on
we take H=1 ): \b (Q_0-2)\K_{\alpha\beta}+\frac{2}{3}{\cal
S}(\bar\partial_{\beta}+2x_{\beta})\bar\partial\cdot
\K_{\alpha}-\frac{1}{3}\theta_{\alpha\beta}\bar\partial\cdot\bar\partial\cdot\K=0\,,\e
By imposing the traceless and divergenceless conditions on the
tensor field $\K_{\alpha\beta}$, which are necessary for UIRs of
dS group, the CI equation (3.1) reduces to \b (Q_{0}-2) {\cal
K}_{\alpha\beta}=0,\,\,\, \mbox{or equivalently} \,\,\, (Q_{2}+4)
{\cal K}_{\alpha\beta}=0\ ,\e we name this CI field equation an
auxiliary field equation.  The relation (2.12) indicates that the
solution of this field equation coincides with the discrete series
normally $\Pi^{\pm}_{2,q=1}$ \cite{16}. Note that in the flat
limit the CI equation $(3.1)$ reduces to CI massless spin-2 wave
equation of order-2 in four dimensional Minkowski space which was
found by Barut and Xu \cite{1}. They have obtained this equation
by varying the coefficients of various term in the standard
equation \cite{1.5}.

Now we want to obtain the solution for this auxiliary field. We
write this solution in general form as \cite {6,18}  \b
\K_{\alpha\beta}=\theta_{\alpha\beta}\phi_1+ {\cal S}\bar
Z_{1\alpha}K_{\beta}+D_{2\alpha}K_{g\beta},\e where the operator
$D_2$ is the generalized gradient defined by
$$D_{2\alpha}K_\beta={\cal S}(\bar\partial_\alpha-x_\alpha)K_\beta,$$ and $ Z_{1} $ is a constant
5-dimensional vector, $ \phi_1 $ is a scalar field, $ K $ and $
K_g $ are two vector fields. The divergenceless
 and transversality conditions together with $ {\cal K}'=0$ result in
\b x\cdot K=0=x\cdot K_g \,\,\,\mbox{and}\,\,\,
2\phi_1+Z_1.K+\bar{\partial}.K_g=0. \e  By substituting $
\K_{\alpha\beta} $ in $(3.2)$ we obtain
  \b \left\{
\ba{rcl}
          (Q_0+4)\phi_1&=&-4Z_1.K,\;\;\;\;\;\;\;\;(I)\\
          Q_1K_\beta&=&0,
          \;\;\;\;\;\;\;\;\;\;\;\;\;\;\;\;\;\;\;(II)\\
          (Q_1+4)K_{g\beta}&=&2(x.Z_1)K_\beta.\;\;\;\;\;(III)
\ea\right.\e Using conditions $ x.K = 0=\bar{\partial}.K $,
Eq.$(3.5-II)$ reduces  to $ (Q_0-2)K_\beta=0$. From this reduced
form and Eq.$(3.5-I)$, we can write \b \phi_1=-\frac{2}{3}Z_1.K,
\;\;\;\mbox{which satisfies}\;\;\;(Q_0-2) \phi_1=0, \e and from
Eq.$(3.4)$, we have \b \bar{\partial}.K_g =\frac{1}{3}Z_1.K.\e
 In continue to our solution, similar to $(3.3)$ we can choose
the following form for the vector field $ K $ (the solution of
$(3.5-II)$) \cite {4,19} \b K_\alpha=\bar Z_{2\alpha}
\phi_2+D_{1\alpha} \phi_3,\e where $D_1=\bar\partial$ and $ Z_2 $
is another 5-dimensional constant vector, $ \phi _2 $ and $ \phi_3
$ are two scalar fields which will be identified later.
Substituting $K$ into $(3.5-II)$ results in \b (Q_0-2)\phi_2 =0,
\e it is clear that $ \phi_2 $ is a massless conformally coupled
scalar field. $\phi_3$ can be written in terms of $\phi_2$
(appendix B) \b \phi_3 =(\ x.Z_2)\phi_2.\e So we can write \b
K_\alpha=\left(\bar Z_{2\alpha}+\ D_{1\alpha}
(x.Z_2)\right)\phi_2,\e
 and
\b \phi_1 =-\frac{2}{3}Z_1.\left(\bar Z_{2}+\ D_{1}(
x.Z_2)\right)\phi_2. \e According to the following identity
(appendix B)\b
(x.Z_1)K_\alpha=\frac{1}{6}(Q_1+4)\left[(x.Z_1)K_\alpha+\frac{1}{9}D_{1\alpha}(Z_1.K)\right],
\e Eq.$(3.5-III)$ leads to \b
K_{g\alpha}=\frac{1}{3}\left[(x.Z_1)K_\alpha+\frac{1}{9}D_{1\alpha}(Z_1.K)\right]+\Lambda_\alpha,\e
where $ x.K_g=0$ and $\bar{\partial}.K_g =\frac{1}{3}Z_1.K$ and
$\Lambda$ is a gauge field with the following conditions \b
 (Q_1+4)\Lambda_\alpha = 0,\,\,\mbox{with}\,\,x\cdot \Lambda = 0\,,\,\, \bar\partial\cdot \Lambda =
 0.\e
 Finally using the Eq.s $(3.11)$, $(3.12)$ and $(3.14)$, we can
rewrite $\K_{\alpha\beta}$ in the following form \b \K_{\alpha
\beta}(x)={\cal D}_{\alpha \beta}(x,\partial,Z_1,Z_2)\phi_2,\e
where ${\cal D}$ is the projector tensor  $$ {\cal
D}(x,\partial,Z_1,Z_2)=\left(-\frac{2}{3}\theta Z_1.+{\cal S}\bar
Z_1+\frac{1}{3}D_2 \left[ \frac{1}{9} D_1 Z_1.+x.Z_1
\right]\right)$$ \b\;\;\;\;\;\;\left( \bar Z_{2}+\ D_{1}\
(x.Z_2)\right).\e
 \setcounter{equation}{0}
\section{Two-point function}

The Wightman two-point function ${\cal W}$ is defined by  \b {\cal
W}_{\alpha\beta \alpha'\beta'}(x,x')=\langle
\Omega|\K_{\alpha\beta}(x)\K_{\alpha'\beta'}(x')|\Omega  \rangle
,\e where $x,x'\in X_H$ and $|\Omega\rangle $ is the Fock vacuum
state. This function which is a solution of the wave Eq.$(3.2)$
with respect to $x$ or $x'$, can be found simply in terms of the
scalar two-point function. It will be shown that this two-point
function can be written as follows $$ {\cal
W}_{\alpha\beta\alpha'\beta'}(x,x')=\Delta_{\alpha\beta\alpha'\beta'}{\cal
W}_c(x,x'),$$ where ${\cal W}_c(x,x')$ is the scalar two-point
function and $\Delta_{\alpha\beta\alpha'\beta'}$ is the bi-tensor
projection operator. We consider the following possibility for the
transverse two-point function
 \b {\cal W}(x,x')=\theta \theta'{\cal W}_0(x,x')+{\cal
S}{\cal S}'\theta.\theta'{\cal W}_{1}(x,x')+D_2D'_2{\cal
W}_g(x,x'),\e where $D_2D'_2=D'_2D_2$ and ${\cal W}_{1}$ and
${\cal W}_{g}$ are transverse bi-vector two-point functions. At
this stage it is shown that calculation of ${\cal W}(x,x')$ could
be initiated from either $x$ or $x'$ without any difference that
means each choices result in the same equation for ${\cal
W}(x,x')$. We first consider the choice $x$. In this case ${\cal
W}(x,x')$ must satisfy the Eq.$(3.2)$, therefor it is easy to show
that: \b \left\{ \ba{rcl}
          (Q_0+4)\theta'{\cal W}_0&=&-4{\cal S}'\theta'.{\cal W}_{1},\;\;\;\;\;\;\;(I)\\
          Q_1{\cal W}_{1}&=&0, \;\;\;\;\;\;\;\;\;\;\;\;\;\;\;\;\;\;\;\;\;(II)\\
          (Q_1+4)D'_2{\cal W}_g&=&2{\cal S}'(x.\theta'){\cal W}_{1}.\;\;\;\;\;(III)
        \ea\right.\e
        Using the condition $\partial.{\cal W}_{1}=0,$  Eq.$(4.3-I)$
leads to \b \theta'{\cal W}_0(x,x')=-\frac{2}{3}{\cal
S}'\theta'.{\cal W}_{1}(x,x').\e The solution of the Eq.$(4.3-II)$
can be written as the combination of two arbitrary bi-scalar
two-point functions ${\cal W}_{2}$ and ${\cal W}_{3}$ in the
following form
$${\cal W}_{1}=\theta.\theta'{\cal W}_{2}+D_1D'_1{\cal W}_{3}.$$
Substituting this in Eq.$(4.3-II)$ and using the divergenceless
condition we obtain
$$D'_1{\cal W}_{3}=\ x.\theta'{\cal W}_{2},\,\,\, \mbox{and}\,\,\,
(Q_0-2){\cal W}_{2}=0.$$ This means that ${\cal W}_{2}$ is the
massless conformally coupled two-point function. Putting ${\cal
W}_{2}\equiv{\cal W}_{c},$ we have \b {\cal
W}_{1}(x,x')=\left(\theta.\theta'
    +\ D_{1}(\ x.\theta')\right){\cal W}_{c}(x,x').\e By using the
following identity
$$(Q_0+4)^{-1}(x.\theta'){\cal W}_{1}=\frac{1}{6}\left[(x.\theta'){\cal W}_{1}+\frac{1}{9}D_1(\theta'.{\cal W}_{1})\right],$$ the Eq.$(4.3-III)$
leads to
 \b D'_2{\cal W}_g(x,x')=\frac{1}{3} {\cal
S'}\left(\frac{1}{9}D_1\theta'.+x.\theta'\right){\cal
W}_{1}(x,x').\e

According to Eq.s $(4.4)$, $(4.5)$ and $(4.6)$ it turns out that
the two-point function can be rewritten in the following form \b
{\cal W}_{\alpha\beta \alpha'\beta'}(x,x')=\Delta_{\alpha\beta
\alpha'\beta'} (x,\partial,x',\partial'){\cal W}_{c}(x,x'), \e
where
$$ \Delta_{\alpha\beta \alpha'\beta'}(x,\partial,x',\partial')=-\frac{2}{3}{\cal S'}\theta
\theta'.\left(\theta.\theta'+\ D_1(\theta'.x)\right)$$
$$ +{\cal S}{\cal S}'\theta.\theta'\left(\theta.\theta'
    +\ D_1(\theta'.x)\right)$$ \b +\frac{1}{3} D_2{\cal
S}'\left(\frac{1}{9}D_1\theta'.+x.\theta'\right)\left(\theta.\theta'
    +\ D_1(\theta'.x)\right).\e

On the other hand with the choice $x'$, the two-point function
$(4.2)$ satisfies  Eq.$(3.2)$ (with respect to $x'$), and
similarly we obtain:
$$ \left\{ \ba{rcl}
          (Q'_0+4)\theta{\cal W}_0&=&-4{\cal S}\theta.{\cal W}_{1},\;\;\;\;\;\;\;\;(I)\\
          Q'_1{\cal W}_{1}&=&0, \;\;\;\;\;\;\;\;\;\;\;\;\;\;\;\;\;\;\;\;(II)\\
          (Q'_1+4)D_2{\cal W}_g&=&2{\cal S}(x'.\theta){\cal W}_{1}.\;\;\;\;\;(III)
        \ea\right.$$
Using the condition $\partial'.{\cal W}_{1}=0,$ we have
$$ \theta{\cal W}_0(x,x')=-\frac{2}{3}{\cal S}\theta.{\cal
W}_{1}(x,x') ,$$ $$ D_2{\cal W}_g(x,x')=\frac{1}{3} {\cal
S}\left(\frac{1}{9}D'_1\theta.+x'.\theta\right){\cal
W}_{1}(x,x'),$$
$$ {\cal W}_{1}(x,x')=\left(\theta'.\theta
    +\ D'_{1}(\ x'.\theta)\right){\cal W}_{c}(x,x'),$$ where the primed operators
act on the primed coordinates only. In this case, the two-point
function can be written in the following form $$ {\cal
W}_{\alpha\beta \alpha'\beta'}(x,x')=\Delta'_{\alpha\beta
\alpha'\beta'} (x,\partial,x',\partial'){\cal W}_{c}(x,x'),
$$ where
$$ \Delta'_{\alpha\beta \alpha'\beta'}(x,\partial,x',\partial')=-\frac{2}{3}{\cal S}\theta
\theta'.\left(\theta.\theta'+\ D'_1(\theta.x')\right)$$
$$ +{\cal S}{\cal S}'\theta.\theta'\left(\theta.\theta'
    +\ D'_1(\theta.x')\right)$$ $$ +\frac{1}{3} D'_2{\cal
S}\left(\frac{1}{9}D'_1\theta.+x'.\theta\right)\left(\theta.\theta'
    +\ D'_1(\theta.x')\right).$$ In a few steps ahead, it is shown that this
equation is non other than  Eq.$(4.8)$.

The conformally coupled scalar field two-point function is
\cite{tag}:
 \b {\cal W}_{c}(x,x')=-\frac{1}{8\pi^2}\left[{\cal P}\frac{1}{1-{\cal Z}(x,x')}-i\pi\epsilon (x^0-x'^0)\delta(1-{\cal Z}(x,x'))\right],
\e where ${\cal P}$ denotes principal part and the geodesic
distance is implicitly defined for $ {\cal{Z}}=-x\cdot x', $ by:
1) $ {\cal{Z}}=\cosh (\sigma )$ if $x$ and $x'$ are time-like
separated,
 2) $ {\cal{Z}}=\cos (\sigma ) $ if
$x$ and $x'$are space-like separated where $\sigma$ is the
distance along the geodesic connecting the points $x$ and $x'$
(note that $ \sigma(x, x')$ can be defined by an unique analytic
extension also when no geodesic connects $x$ and $x'$), and also \b \epsilon (x^0-x'^0)=\left\{ \ba{rcl} 1&x^0>x'^0 ,\\
0&x^0=x'^0 ,\\ -1&x^0<x'^0.\\ \ea\right.\e

Eq.s $(4.4)$, $(4.5)$, $(4.6)$ and $(4.9)$ after relatively simple
and straightforward calculations can be written as (appendix A):
\b\theta'_{\alpha'\beta'}{\cal W}_0(x,x') =-\frac{2}{3}{\cal
S}'\left[2\theta'_{\alpha'\beta'}
+(x.\theta'_{\alpha'})(x.\theta'_{\beta'})(2+{\cal{Z}}\frac{d}{d{\cal{Z}}})\right]{\cal
W}_{c}({\cal{Z}}),\e \b{\cal W}_{1\beta
\beta'}(x,x')=\left[-(x'.\theta_{\beta})(x.\theta'_{\beta'})\frac{d
}{d{\cal{Z}}} +2(\theta_{\beta}.\theta'_{\beta'})\right]{\cal
W}_{c}({\cal{Z}}),\e
$$D_{2\alpha}D'_{2\alpha'}{\cal
W}_{g\beta\beta'}(x,x')=\frac{1}{27(1-{\cal{Z}}^2)^2}{\cal S}{\cal
S}'
\left[\theta_{\alpha\beta}\theta'_{\alpha'\beta'}(1-{\cal{Z}}^2)^2(2{\cal{Z}}\frac{d}{d{\cal{Z}}})\right.$$
$$+(x'.\theta_{\alpha})
(x'.\theta_{\beta})(x.\theta'_{\alpha'})(x.\theta'_{\beta'})((26-14{\cal{Z}}^2)+(58{\cal{Z}}-34{\cal{Z}}^3)\frac{d}{d{\cal{Z}}})$$
$$+\theta'_{\alpha'\beta'}(x'.\theta_{\alpha})
(x'.\theta_{\beta})(1-{\cal{Z}}^2)(4+8{\cal{Z}}\frac{d}{d{\cal{Z}}})$$$$+\theta_{\alpha\beta}
(x.\theta'_{\alpha'})(x.\theta'_{\beta'})(1-{\cal{Z}}^2)((22-20{\cal{Z}}^2)+(14{\cal{Z}}-10{\cal{Z}}^3)\frac{d}{d{\cal{Z}}})$$
$$+(\theta_{\alpha}.\theta'_{\alpha'})(\theta_{\beta}.\theta'_{\beta'})(1-{\cal{Z}}^2)^2(22+2{\cal{Z}}\frac{d}{d{\cal{Z}}})$$
\b\left.-(\theta_{\alpha}.\theta'_{\alpha'})(x.\theta'_{\beta'})(x'.\theta_{\beta})(1-{\cal{Z}}^2)(8{\cal{Z}}+(48-32{\cal{Z}}^2)\frac{d}{d{\cal{Z}}})
\right]{\cal W}_{c}({\cal{Z}}),\e where $$ (Q_0-2){\cal
W}_{c}({\cal{Z}})=0.$$ Now we are in a position to write the final
form of the two-point function in ambient space. Substituting
Eq.s$(4.11)$, $(4.12)$ and $(4.13)$ in $(4.2)$ yields
$$ {\cal W}_{\alpha\beta
\alpha'\beta'}(x,x')=\frac{1}{27}{\cal S}{\cal
S}'\left[\theta_{\alpha\beta}\theta'_{\alpha'\beta'}f_1({\cal{Z}})+(\theta_{\alpha}.\theta'_{\alpha'})(\theta_{\beta}.\theta'_{\beta'})f_2({\cal{Z}})\right.$$
$$+\theta'_{\alpha'\beta'}(x'.\theta_{\alpha})
(x'.\theta_{\beta})f_3({\cal{Z}})+(x'.\theta_{\alpha})
(x'.\theta_{\beta})(x.\theta'_{\alpha'})(x.\theta'_{\beta'})f_4({\cal{Z}})+(\theta_{\alpha}.\theta'_{\alpha'})(x.\theta'_{\beta'})(x'.\theta_{\beta})f_5({\cal{Z}})
$$
\b\label{2}\left.
+\theta_{\alpha\beta}(x.\theta'_{\alpha'})(x.\theta'_{\beta'})
f_6({\cal{Z}})\right]{\cal W}_{c}({\cal{Z}}).\e where

$$f_1({\cal{Z}})=(-18+2{\cal{Z}}\frac{d}{d{\cal{Z}}}),\,\,\,f_2({\cal{Z}})=(76+2{\cal{Z}}\frac{d}{d{\cal{Z}}}),$$
$$f_3({\cal{Z}})=(1-{\cal{Z}}^2)^{-1}\left(4+8{\cal{Z}}\frac{d}{d{\cal{Z}}}\right),$$
$$f_4({\cal{Z}})=(1-{\cal{Z}}^2)^{-2}\left((26-14{\cal{Z}}^2)+(58{\cal{Z}}-34{\cal{Z}}^3)\frac{d}{d{\cal{Z}}}\right),$$
$$f_5({\cal{Z}})=-(1-{\cal{Z}}^2)^{-1}\left(8{\cal{Z}}+(75-59{\cal{Z}}^2)\frac{d}{d{\cal{Z}}}\right),$$
$$f_6({\cal{Z}})=(1-{\cal{Z}}^2)^{-1}\left( (4-2{\cal{Z}}^2)+(5{\cal{Z}}-{\cal{Z}}^3)\frac{d}{d{\cal{Z}}}\right).$$
Eq.$(4.14)$ is the explicit form of the two-point function in 5
dimensional flat space (ambient space). This equation satisfies
the traceless and divergenceless conditions:
$$\bar{\partial}.{\cal
W}=\bar{\partial'}.{\cal W}=0,\;\;\;\mbox{and}\;\;\;{\cal
W}_{\alpha\beta \alpha'}^{\;\;\;\;\;\alpha'}(x,x')={\cal
W}^{\alpha}_{\;\;\alpha \alpha'\beta'}(x,x')=0.$$   The two-point
function $(4.14)$ is obviously dS-invariant.\\ Now it is
straightforward to translate this two point function on 4
dimensional de Sitter hyperboloid (intrinsic coordinate)as follows
(appendix C)
$$
Q_{\mu\nu\mu'\nu'}(X,X')=\frac{1}{27}{\cal S}{\cal
S}'\left[\;g_{\mu\nu}g'_{\mu'\nu'}\;\frac{f_1}{(1-{\cal{Z}}^2)^2}
+g_{\mu\mu'}g_{\nu\nu'}\;\frac{f_2}{(1-{\cal{Z}}^2)^2}\right.$$
$$+\;g'_{\mu'\nu'}n_{\mu}
n_{\nu}\;\frac{f_3}{1-{\cal{Z}}^2}+g_{\mu\mu'}n_\nu
n_{\nu'}\left(\frac{2({\cal{Z}}-1)f_2}{(1-{\cal{Z}}^2)^2}
+\frac{f_5}{1-{\cal{Z}}^2} \right)
$$
\b\left.+n_{\mu}
n_{\nu}n_{\mu'}n_{\nu'}\left(\frac{f_2}{(1+{\cal{Z}})^2}-\frac{f_5}{1+{\cal{Z}}}+f_4
\right)+g_{\mu\nu}n_{\mu'}n_{\nu'}\;\frac{f_6}{1-{\cal{Z}}^2}
\right]{\cal W}_{c}({\cal{Z}}).\e The two-point function $(4.15)$
is clearly dS-invariant.
\section{Conclusion}
Conformal invariance is the important common property for all
equations of massless fields; for example massless vector field
(photon) satisfies Maxwell equation which is CI. So, it seems to
be important to obtain a CI equation for the massless spin-2
field. Einstein's equation in its linear form is often interpreted
as the equation for spin-2 massless field (graviton) in a fixed
background metric. Einstein's classical theory of gravitation is
not CI, thus could not be considered as a comprehensive universal
theory of gravitational fields. \\ In the Dirac's six-cone
formalism by setting $n=1$ in (\ref{di}) we obtained the CI wave
equation for $\K_{\alpha\beta}$ which transformed according to the
UIR $(\Pi^{\pm}_{2,1})$ in dS space \cite{1}. Minkowskian limit
($H\rightarrow 0$) of this CI equation coincides to what reported
in Ref. \cite{1.5}. In this paper we considered the solution of
this equation. Imposing the conditions divergencelessness and
transversality on this field we obtained
$(Q_2+4)\K_{\alpha\beta}=0$. On the other hand it has been shown
that rank-2 tensor field which is the UIR of the dS and conformal
groups satisfies following equation \cite{03} $$
(Q_2+4)^2(Q_2+6)\K_{\alpha\beta}=0,$$ this 6-order differential
equation in its non-linear form leads to $R^3$-gravity in dS
space. The solutions of the second part of the above equation
($(Q_2+6)\K_{\alpha\beta}=0$) has been considered previously. In
this paper we studied the solutions of the
$(Q_2+4)\K_{\alpha\beta}=0$. We find the solution for this CI
field equation in terms of the massless conformally coupled scalar
field. Related two-point function was also written in terms of the
massless conformally coupled two-point function. These two-point
function is then dS-invariant. This method may pay the road in
considering the field equation of the higher order derivative
theories in dS space especially $R^3$ theory of gravity.

\noindent {\bf{Acknowledgement}}: One of us H.P would like to
thank A. Pourmajidi for her interest in this work.

 \setcounter{equation}{0}
\begin{appendix}
\section{Some useful relations}

 In this appendix, some useful relations are given: \b Q_1 D_1 =D_1
Q_0 ,\e  \b
(Q_{0}-2)x_{\alpha}=x_{\alpha}Q_{0}-6x_{\alpha}-2\bar\partial_{\alpha},
\e \b
\bar\partial_{\alpha}(Q_{0}-2)=Q_{0}\bar\partial_{\alpha}-8\bar\partial_{\alpha}-
2Q_{0}x_{\alpha} - 8x_{\alpha},\e \b
x_{\alpha}Q_{0}(Q_{0}-2)=(Q_{0}-2)(Q_{0}x_{\alpha}+4x_{\alpha}+4\bar\partial
_{\alpha}), \e  \b [Q_{0}Q_{2},Q_{2}Q_{0}]{\cal
K}_{\alpha\beta}=4{\cal
S}(x_{\alpha}-\bar\partial_{\alpha})\bar\partial .{\cal
K}_{\beta}.\e  Following relations become important in deriving
two-point function \b \bar{\partial}_\alpha
f({\cal{Z}})=-(x'.\theta_{\alpha})\frac{d
f(\cal{Z})}{d{\cal{Z}}},\e
\b\theta^{\alpha\beta}\theta'_{\alpha\beta}=\theta..\theta'=3+{\cal{Z}}^2,
\e \b (x.\theta'_{\alpha'})(x.\theta'^{\alpha'})={\cal{Z}}^2-1,\e
\b
(x.\theta'_{\alpha})(x'.\theta^{\alpha})={\cal{Z}}(1-{\cal{Z}}^2),\e
\b\bar{\partial}_\alpha(x.\theta'_{\beta'})=\theta_{\alpha}.\theta'_{\beta'},\e
\b\bar{\partial}_\alpha(x'.\theta_{\beta})=x_\beta(x'.\theta_{\alpha})-{\cal{Z}}\theta_{\alpha\beta},\e
\b\bar{\partial}_\alpha(\theta_{\beta}.\theta'_{\beta'})=x_\beta(\theta_{\alpha}.\theta'_{\beta'})+
\theta_{\alpha\beta}(x.\theta'_{\beta'}),\e
\b\theta'^{\beta}_{\alpha'}(x'.\theta_{\beta})=-{\cal{Z}}(x.\theta'_{\alpha'}),\e
\b\theta'^{\gamma}_{\alpha'}(\theta_{\gamma}.\theta'_{\beta'})=\theta'_{\alpha'\beta'}+(x.\theta'_{\alpha'})(x.\theta'_{\beta'}),\e
\b Q_0f({\cal{Z}})=(1-{\cal{Z}}^2)\frac{d^2
f(\cal{Z})}{d{\cal{Z}}^2}-4{\cal{Z}}\frac{d
f(\cal{Z})}{d{\cal{Z}}}.\e

\setcounter{equation}{0}
\section{Some details on equations (3.10) and (3.13)}
Substituting K into (3.5-II) results in $
Q_0{\phi}_3=2x.{\cal{Z}}_2{\phi}_2.$ Imposing the divergenceless
condition, we get $
2x.{\cal{Z}}_2{\phi}_2=-{\cal{Z}}_2.\bar\partial{\phi}_2.$ Using
the above relations and (3.9) and (A.2) one can obtain (3.10).

By using $(2.13)$ it is easy to show that \b
D_1(Z_1.K)=\frac{1}{6}(Q_1+4)[D_1(Z_1.K)],\e \b
x(Z_1.K)=\frac{1}{6}(Q_1+4)[x(Z_1.K)], \e \b
Z_1.\bar{\partial}K=\frac{1}{6}(Q_1+4)[Z_1.\bar{\partial}K-\frac{1}{3}D_1(Z_1.K)],\e
\b (Q_1+4)[(x.Z_1)K]=2[x(Z_1.K)-Z_1.\bar{\partial}K].\e The
conditions $ x.K=\bar{\partial}.K=0, $ and $(Q_0-2)K=0, $ are used
to obtain the above equations.

Substituting Eq.s $(B.2)$ and $(B.3)$ in $(B.4)$ we have \b
(Q_1+4)[(x.Z_1)K]=\frac{1}{3}(Q_1+4)\left[\frac{1}{3}D_1(Z_1.K)+x(Z_1.K)-Z_1.\bar{\partial}K\right],\e
or \b
(x.Z_1)K=\frac{1}{3}\left[\frac{1}{3}D_1(Z_1.K)+x(Z_1.K)-Z_1.\bar{\partial}K\right].\e
Finally according Eq.s $(B.1)$ and $(B.4)$, we obtain
\b(x.Z_1)K=\frac{1}{6}(Q_1+4)\left[\frac{1}{9}D_1(Z_1.K)+(x.Z_1)K
\right].\e This automatically leads to  Eq.$(3.13)$.

\setcounter{equation}{0}
\section{Relation between the ambient space notation and the intrinsic coordinates}

In order to compare our results with the work of the other authors
\cite{hiko1,28}, the relation between the ambient space notation
and the intrinsic coordinates is studied in the final stage. In
order to translate the relations into the ambient coordinates, we
use the fact that the ``intrinsic'' field $h_{\mu\nu}(X)$ is
locally determined by the ``transverse'' tensor field
$\K_{\alpha\beta}(x)$ through \b\lbl{passage}
h_{\mu\nu}(X)=\frac{\partial x^{\alpha}}{\partial
X^{\mu}}\frac{\partial x^{\beta}}{\partial
X^{\nu}}\K_{\alpha\beta}(x(X)). \e  In the same way one can show
that the transverse projector $\theta$ is the only symmetric and
transverse tensor which is linked to the dS metric $g_{\mu\nu}$:
$$g_{\mu\nu}=\frac{\partial x^{\alpha}}{\partial
X^{\mu}}\frac{\partial x^{\beta}} {\partial
X^{\nu}}\,\theta_{\alpha\beta}.$$ Covariant derivatives acting on
a symmetric, second rank tensor are transformed according to \b
\nabla_{\rho}\nabla_{\lambda}h_{\mu\nu}= \frac{\partial
x^{\alpha}}{\partial X^{\rho}}\frac{\partial x^{\beta}}{\partial
X^{\lambda}}\frac{\partial x^{\gamma}}{\partial
X^{\mu}}\frac{\partial x^{\delta}}{\partial
X^{\nu}}\mbox{Trpr}\bar
\partial_\alpha \mbox{Trpr}\bar\partial_\beta \K_{\gamma\delta}.
\e The transverse projection (Trpr) defined by $$
(\mbox{Trpr}\K)_{\alpha\beta}=\theta^\gamma_\alpha
\theta^\delta_\beta \K_{\gamma\delta},$$ guarantees the
transversality in each index. For example we have \cite{6}
$$\nabla_{\rho}\nabla_{\lambda}h_{\mu\nu}= \frac{\partial
x^{\alpha}}{\partial X^{\rho}}\frac{\partial x^{\beta}}{\partial
X^{\lambda}}\frac{\partial x^{\gamma}}{\partial
X^{\mu}}\frac{\partial x^{\delta}}{\partial X^{\nu}}[
\bar\partial_\alpha(\bar\partial_\beta \K_{\gamma\delta}-
x_{\gamma}\K_{\beta\delta}- x_{\delta}\K_{\beta\gamma})$$ $$ -
x_\beta(\bar\partial_\alpha \K_{\gamma\delta}-
x_{\gamma}\K_{\alpha\delta}- x_{\delta}\K_{\alpha\gamma}
)-x_\gamma(\bar\partial_\beta \K_{\alpha\delta}-
x_{\alpha}\K_{\beta\delta}- x_{\delta}\K_{\beta\alpha})$$ \b
-x_\delta(\bar\partial_\beta \K_{\gamma\alpha}-
x_{\gamma}\K_{\beta\alpha}- x_{\alpha}\K_{\beta\gamma})].\e By
contraction of the covariant derivatives, {\it i.e.}
$\nabla_{\rho}\nabla^{\rho}$, the d'Alambertian operator becomes:
\b \Box_{H}
h_{\mu\nu}=g^{\lambda\rho}\nabla_{\lambda}\nabla_{\rho}h_{\mu\nu}=\frac{\partial
x^{\alpha}}{\partial X^{\mu}}\frac{\partial x^{\beta}}{\partial
X^{\nu}}([\bar \partial_\gamma \bar \partial^\gamma
-2]\K_{\alpha\beta}-2{\cal S}x_\alpha( \bar \partial.\K)_\beta
+2x_\alpha x_\beta \K'),\e other relations can be found by this
way.

As mentioned in \cite{6}, any maximally symmetric bi-tensor are
functions of two points $(x,x')$ and behave like tensors under
coordinate transformations at each points and they can be
expressed as a sum of products of three basic tensors. The
coefficients in this expansion are functions of the geodesic
distance $\sigma(x, x')$. In this sense, these fundamental tensors
form a complete set and they can be obtained by differentiating
the geodesic distance:
$$n_\mu = \nabla_\mu \sigma(x, x')\;\;\;,\;\;\; n_{\mu'} = \nabla_{\mu'} \sigma(x,
x'), $$and the parallel propagator
$$g_{\mu\nu'}=-c^{-1}({\cal{Z}})\nabla_{\mu}n_{\nu'}+n_\mu n_{\nu'}.$$
 The basic bi-tensors in
ambient space notations are found through
$$ \bar{\partial}_\alpha \sigma(x,x')\;\;\;,\;\;\;\bar{\partial}'_{\beta'}
\sigma(x,x')\;\;\;,\;\;\;\theta_\alpha .\theta'_{\beta'},$$
restricted to the hyperboloid by
$$ {\cal{T}}_{\mu\nu'}=\frac{\partial x^\alpha}{\partial
X^\mu}\frac{\partial x'^{\beta'}}{\partial
X'^{\nu'}}T_{\alpha\beta'}.$$

For $ {\cal{Z}}=\cos(\sigma), $ one can find
$$n_\mu=\frac{\partial x^\alpha}{\partial X^\mu}\bar{\partial}_\alpha \sigma(x,x')=
\frac{\partial x^\alpha}{\partial X^\mu} \frac{(x' \cdot
\theta_\alpha)}{\sqrt{1-{\cal{Z}}^2}},\;\; n_{\nu'}=\frac{\partial
x'^{\beta'}}{\partial X'^{\nu'}}\bar{\partial}_{\beta'}
\sigma(x,x') =\frac{\partial x'^{\beta'}}{\partial X'^{\nu'}}
\frac{(x\cdot\theta'_{\beta'})}{\sqrt{1-{\cal{Z}}^2}},$$
$$\nabla_\mu n_{\nu'}=\frac{\partial x^\alpha}{\partial
X^\mu}\frac{\partial x'^{\beta'}}{\partial
X'^{\nu'}}\theta^\varrho_\alpha
\theta'^{\gamma'}_{\beta'}\bar{\partial}_\varrho\bar{\partial}_{\gamma'}
\sigma(x, x')=c({\cal{Z}})[n_\mu n_{\nu'}{\cal{Z}}-\frac{\partial
x^\alpha}{\partial X^\mu}\frac{\partial x'^{\beta'}}{\partial
X^{\nu'}}\theta_\alpha \cdot\theta'_{\beta'}],$$ with $
c^{-1}({\cal{Z}})=-\frac{1}{\sqrt{1-{\cal{Z}}^2}}.$  For $
{\cal{Z}}=\cosh (\sigma), $  $ n_\mu $ and $ n_\nu$ are multiplied
by $i$ and $ c({\cal{Z}}) $ becomes
$-\frac{i}{\sqrt{1-{\cal{Z}}^2}}.$ In both cases we have
$$ g_{\mu\nu'}+({\cal{Z}}-1)n_\mu n_{\nu'}=\frac{\partial x^\alpha}{\partial
X^\mu}\frac{\partial x'^{\beta'}}{\partial X'^{\nu'}}\theta_\alpha
\cdot\theta'_{\beta'} .$$ and the two-point functions are related
through $$ Q_{\mu\nu\mu'\nu'}= \frac{\partial x^\alpha}{\partial
X^\mu} \frac{\partial x^\beta}{\partial X^\nu} \frac{\partial
x'^{\alpha'}}{\partial X'^{\mu'}} \frac{\partial
x'^{\beta'}}{\partial
X'^{\nu'}}{\cal{W}}_{\alpha\beta\alpha'\beta'}.$$

\end{appendix}

\end{document}